\documentclass[runningheads]{llncs}
\usepackage[T1]{fontenc}
\usepackage{hyperref}
\usepackage{graphicx}
\begin{document}
\title{Enabling Formal Verification in a Common Model of Cognition}
%
%
\author{Parth Ganeriwala\inst{1} \and
Miki Matsumuro\inst{2} \thanks{Present address: Honda Research Institute Japan Co., Ltd., Wako, Japan (\email{miki.matsumuro@jp.honda-ri.com})} \and\\
Frank E. Ritter\inst{2}  \and Siddhartha Bhattacharyya\inst{1}}
\authorrunning{Ganeriwala et al.}
%
\institute{College of COES, Florida Institute of Technology, Melbourne, FL \and
College of IST, The Pennsylvania State University, Centre County, PA\\
\email{\{pganeriwala2022,sbhattacharyya\}@fit.edu},
\email{miki.matsumuro@gmail.com},
\email{frank.ritter@psu.edu}}

\maketitle              
\begin{abstract}
The development of accurate and efficient cognitive models remains a central challenge in cognitive science and artificial intelligence. This paper introduces a novel methodology that integrates formal verification techniques into cognitive model development, using the Common Model of Cognition (CMoC) as a unifying framework. By translating cognitive models into state machine representations, we enable automated analysis through formal verification tools such as \textit{nuXmv}. We demonstrate this approach using three toy models—ranging from procedural execution, memory retrieval, and problem-solving—which are developed in both ACT-R and Soar architectures. These models are transformed into state transition systems to verify properties such as correctness (i.e., task completion) and haltability (i.e., guaranteed termination). Our method offers a pathway to increasing confidence in model behavior and reveals insights into the minimal constructs required to represent and translate cognitive architectures into formal verification environments. This work highlights the viability of CMoC as a state machine framework suitable for formal reasoning, and offers a pathway toward rapid, accurate, and architecture-neutral cognitive model development. We conclude by outlining how these foundations can support the creation of a high-level modeling language and compiler that bridges symbolic cognitive architectures with formal methods from computer science.



\end{abstract}

\section{Introduction}

Cognitive models offer a powerful means of simulating and understanding human thought processes, with applications in psychology, human–computer interaction, and autonomous systems. Architectures such as ACT-R \cite{anderson2013adaptive} and Soar \cite{laird2019soar} have enabled the development of numerous task models, capturing phenomena ranging from memory retrieval to problem solving. However, while these models often succeed in mimicking behavioral outcomes, they typically lack formal guarantees regarding their internal logic, completeness, and termination. This gap between empirical fidelity and logical soundness presents an important challenge. As cognitive models become more complex and are applied in critical settings—such as adaptive training systems, intelligent agents, and cognitive tutoring—assessing their correctness and reliability becomes increasingly important, especially as they are deployed in critical settings. Traditional validation techniques, including behavioral alignment and expert inspection, do not always scale or catch subtle implementation flaws.

To address this limitation, we explore the use of formal verification methods from computer science, particularly model checking, as a complementary tool for cognitive modeling. Our aim is to move toward a framework where cognitive models can be analyzed not only for plausibility but also for logical consistency and correctness. Central to our approach is the Common Model of Cognition (CMoC) \cite{lairdLR17}, a proposed standard architecture that captures shared structural principles across multiple cognitive systems. The CMoC provides an abstract, modular view of cognition—including working memory buffers, long-term memory systems, and a central procedural controller—that lends itself naturally to formal representation. By treating CMoC as a high-level design language, we can map cognitive behaviors into state machines suitable for formal analysis.

\begin{figure}[hb!tp]
    \centering
    \includegraphics[width=\linewidth]{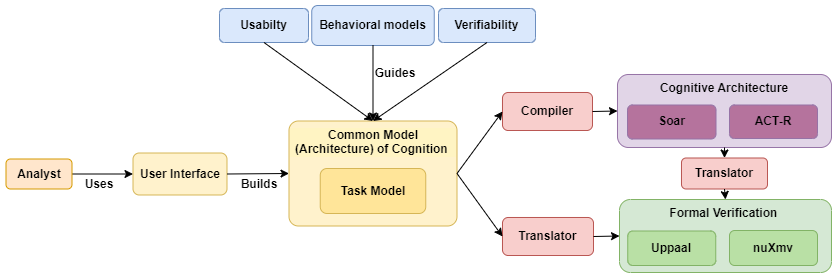}
    \caption{Overview of the verification process and compiler approach.}
    \label{fig:enter-label}
\end{figure}

In this paper, we investigate this approach through three illustrative cognitive tasks that span procedural control, memory retrieval, and problem solving. These tasks are implemented in both ACT-R and Soar to highlight architectural differences and commonalities. We then translate these models into formal state-transition systems and analyze them using the nuXmv model checker \cite{cavada2014}, focusing on properties such as haltability and goal satisfaction. Our broader goal is to lay the groundwork for a toolchain that supports verifiable cognitive modeling across architectures. This includes defining the minimal constructs needed for formal translation, identifying architectural features that align well with verification tools, and exploring the potential for automated compilation from cognitive specifications to formal models. Figure~\ref{fig:enter-label} shows an overview of our verification pipeline and compiler-based translation process.

The remainder of this paper is organized as follows. We begin by summarizing the related work in this domain and then outlining the theoretical background of the CMoC, ACT-R, and Soar. We then describe our three case models and their formalization. Finally, we present our findings and discuss implications for the future development of high-assurance cognitive modeling frameworks.

\section{Related Work}
Cognitive architectures like ACT-R and Soar have been targets for applying formal verification methods (e.g., model checking, formal semantic modeling) to ensure the correctness of cognitive models \cite{11014784}. For instance, Langenfeld et al. \cite{langenfeld2019formal} present a formal operational model of the ACT-R architecture along with a translation of ACT-R cognitive models into timed automata, enabling exhaustive analysis of model behavior. Using this approach, they could automatically check ACT-R models for defects (such as deadlocks or incorrect rule implementations) and verify properties like expected chunk presence or timing feasibility, which would be tedious to find by simulation alone. This work builds on prior formal semantics for ACT-R (e.g., Albrecht \& Westphal \cite{albrecht2014analyzing}) and demonstrates that model checking techniques can scale to non-trivial ACT-R models. Other researchers have explored specific formal properties; for example, Gall and Frühwirth \cite{gall2017decidable} encoded ACT-R in constraint-handling rules to verify confluence (i.e. consistency of outcomes despite rule ordering) in ACT-R models. These efforts highlight a clear connection between cognitive science models and formal methods: by treating cognitive architectures as formal systems (e.g., state machines or logical frameworks), one can apply verification tools to guarantee that cognitive models faithfully implement the intended psychological theory.  

Early work by Macklem and Mili \cite{macklem2006formal} introduced a specification language for cognitive models and algorithms to automatically check a model’s correctness against its specifications. This was among the first attempts to bring rigorous software verification practices to cognitive modeling, underlining the need for modular, reusable model specifications and the ability to prove that a cognitive model meets certain requirements (both in terms of task performance and human-like behavior). 

Beyond ACT-R, other architectures have seen similar efforts. In the Soar architecture, recent research has integrated formal verification with cognitive modeling for autonomous agents. Ganeriwala et al. (2025) describe a framework where agent decision logic is modeled in Soar (including its symbolic rules and reinforcement learning components) and then automatically translated into the nuXmv model checker to verify safety and operational correctness properties \cite{ganeriwala2025systems}. This combination of systems modeling, simulation, and model checking ensures that the Soar-based autonomous agent behaves correctly under all considered scenarios, marrying cognitive architecture design with formal verification to improve reliability. Similarly, Ragni et al. \cite{ragni2018towards} argued for a formal foundation for cognitive architectures in general, using formal representations to compare and evaluate different architectures. All of these works illustrate how formal methods (such as symbolic model analysis, temporal logic model checking, or theorem proving) can be applied to cognitive architectures to rigorously verify models of human cognition.

\section{Preliminaries}
\subsection{Common Model of Cognition}
The Common Model of Cognition (CMoC) – originally proposed as the “Standard Model of the Mind” by Laird, Lebiere, and Rosenbloom \cite{lairdLR17} – is a high-level consensus framework capturing the core cognitive modules and their interactions common to many cognitive architectures. In essence, the CMoC distills decades of cognitive architecture research (spanning ACT-R, Soar, SIGMA, etc.) into a unified model of the mind’s fixed structure \cite{romero2018cogarch}. It specifies key processing modules (perception, working memory, procedural memory, action, etc.) and the information flow between them as a blueprint for both understanding human cognition and designing cognitive systems.

A number of studies have applied and evaluated the CMoC in various contexts, often integrating it with computational tools or data to test its validity. For example, Stocco et al. \cite{stocco2021analysis} conducted a neuroimaging-based evaluation of the CMoC by mapping its components to brain regions and comparing its predicted brain connectivity patterns against human fMRI data across multiple tasks. Using dynamic causal modeling and a Bayesian model comparison, they found that a network architecture based on the Common Model "vastly outperforms all other architectures" (several alternative theoretical brain architectures) in explaining the fMRI data. This provides strong empirical support that the CMoC’s functionally defined modules and pathways correspond to the organization of human cognition in the brain, highlighting a bridge between a formal cognitive architecture and biological data. 

Researchers have also worked on integrating additional cognitive facets into the Common Model and formalizing its structure. For instance, Larue et al. \cite{larue2018emotion} explored how emotion could be incorporated into the CMoC, identifying the functional points where emotion processes interface with the standard cognitive modules. This study outlined how various models of emotion can be aligned with the CMoC, acknowledging that human-like AI will require motivational and emotional states within the cognitive architecture. In a similar vein, other extensions have been proposed: e.g., adding metacognitive capabilities (self-monitoring and control) on top of the CMoC’s processes and evaluating how this could be realized in a unified model \cite{kralik2018metacognition}, or examining how natural language processing fits into the Common Model’s structure \cite{kelly2018language}. These efforts apply the CMoC as a baseline and then integrate new components or constraints, often via computational implementation, to test the model’s completeness and adaptability. 

Notably, there have also been moves to use formal methods to specify and verify the CMoC itself. Romero \cite{romero2018cogarch} proposed CogArch-ADL, an architecture description language (ADL) based on $\pi$-calculus, to formally describe cognitive architectures in a CMoC-compliant way. By extending a formal ADL, Romero’s approach allows researchers to rigorously specify the modules and connections of a CMoC-based architecture and then verify properties like architectural consistency or equivalence between different cognitive architectures. This kind of work ensures that as new cognitive architectures are developed or existing ones are compared, they can be checked against the Common Model’s formal specification, thereby bridging cognitive architecture theory with formal verification at the architectural level. Such formal descriptions make it possible to prove, for example, that a given implementation of a cognitive architecture conforms to the CMoC’s structural requirements or to model-check high-level cognitive workflows for logical soundness within the CMoC framework.

\subsection{ACT-R}
ACT-R is a cognitive architecture and a theory of simulating and understanding human cognition \cite{anderson2013adaptive,ritter2019act}. Its theory is embodied in the ACT-R software, through which we can construct models that can store, retrieve, and process knowledge, as well as explain and predict performance \cite{bothell2017act}.
There are currently two main kinds of knowledge representations in ACT-R, and they are declarative knowledge and procedural knowledge. Declarative knowledge consists of chunks of memory (e.g., apple is a kind of fruit), while procedural knowledge performs basic operations, moves data among buffers, and identifies the next instructions to be executed (e.g., if the preparation to submit your answer is completed, then click the submit button). When the model is driving a bus from a first-person perspective, these pieces of information will contain information such as what visual items are presented to look at and what tasks to do next \cite{schwartz2020drive}.

\subsection{Soar}
Soar is a general cognitive architecture that provides a computational infrastructure that resembles the cognitive capabilities exhibited by a human. Soar implements knowledge-intensive reasoning that enables the execution of rules based on the context. It also has the capability to integrate learning into the agent using chunking or reinforcement learning. Soar has its origins in the work done by Newell and Simon \cite{newell1972human} from the late
1950s through the mid-1970s, also inspired by the ”General Problem Solver” created by Ernst and Newell \cite{newell1961gps}. While ACT-R was designed to model human behavior, Soar was inspired by it. Soar’s general computing concept is based on objectives, problem spaces, states and operators \cite{laird2019soar,newell1994unified} . Soar encompasses multiple memory constructs (e.g., semantic, episodic, etc.) and learning mechanisms (e.g., reinforcement, chunking, etc.) and is a programmable architecture with an embedded theory. This enables executing Soar models on embedded system platforms and studying the design problem through rapid prototyping and simulation.
\subsection{nuXmv} 
nuXmv is a symbolic model checker. It builds on and extends NuSMV. It implements verification for finite and infinite state synchronous transition systems. For finite-state systems, it complements NuSMV's basic verification techniques with a family of new state-of-the-art verification algorithms. For infinite-state systems, it extends the NuSMV language with new data types, namely integers and reals, and it provides advanced SMT-based model checking techniques. nuXmv implements SMT-based model checking techniques \cite{cavada2014}.

\section{Initial Development}
We developed and analyzed three cognitive models to evaluate our framework: a simple procedural task (Hungry-Thirsty), a memory retrieval task (Fan Effect), and a problem-solving task (Water Jug). Each was implemented in ACT-R and Soar, then translated into state machine representations to enable formal verification using the CMoC. The corresponding source code and model files are available on \href{https://github.com/ParthGaneriwala/FMinCMoC.git}{GitHub}.

\subsection{Development Case 1: Hungry-Thirsty Model}
The Hungry-Thirsty model \cite{httut} illustrates a simple decision-making task and serves as a canonical example for teaching and comparing production-based behavior across architectures. The agent is either hungry, thirsty, or both. If hungry, it eats; if thirsty, it drinks. If both conditions are true, eating takes priority over drinking.
\paragraph{\textbf{Soar model.}}
In Soar, the model uses two attributes—\texttt{hungry} and \texttt{thirsty}—each with binary values \texttt{yes} or \texttt{no}. Operators \texttt{EAT} and \texttt{DRINK} transition the agent’s state by satisfying hunger or thirst respectively. Applying \texttt{EAT} sets \texttt{hungry = no} while leaving thirst unchanged. Likewise, \texttt{DRINK} updates only the thirst state. Production rules define operator applicability based on current attributes, and the system halts when both drives are satisfied.

\paragraph{\textbf{ACT-R model.}}
The ACT-R implementation stores the agent’s state in the goal buffer. Three production rules control the behavior: one each for \texttt{eat}, \texttt{drink}, and \texttt{stop}. Rule selection is based on utility values; eating is prioritized by assigning it a higher utility than drinking. When neither hunger nor thirst is present, the stop rule halts the model.

\subsection{Development Case 2: Fan Model}
The Fan Effect task in the ACT--R tutorial (Unit 5 \url{http://act-r.psy.cmu.edu/software/}) demonstrates how memory retrieval is influenced by associative complexity. The model learns sentences associating people with locations (e.g., "A hippie is in the park"). Later, it must determine whether new probe sentences match learned pairs. Reaction time increases with the number of associations—a phenomenon the model aims to replicate.
\paragraph{\textbf{Soar model.}}
In Soar, learned pairs are explicitly stored in semantic memory. Probes activate retrieval rules based on working memory cues. Retrieval latency is simulated by requiring multiple decision cycles when several memories match, reflecting cognitive load.

\paragraph{\textbf{ACT-R model.}}
The ACT-R version follows the tutorial design. Learned associations reside in declarative memory, and probes are encoded in the imaginal buffer. ACT-R's activation-based retrieval naturally captures the fan effect: when multiple chunks match a cue, spreading activation is diluted, increasing retrieval time. The model’s timing behavior closely reflects empirical human data.

\subsection{Development Case 3: Water Jug Model}
This task involves reaching a target volume of water (e.g., 2 liters) using two jugs with different capacities (3 and 5 liters). It is a classic benchmark for problem solving through state-space exploration and the Einstellung (problem solving set) effect.
\paragraph{\textbf{Soar model.}}
Soar solves the task using a defined problem space and operators: \texttt{fill}, \texttt{empty}, and \texttt{pour}. Each operator has explicit preconditions and modifies the agent’s working memory to reflect the current jug state. To avoid revisiting previous states, Soar uses episodic memory to track transitions, ensuring efficient path traversal.
\paragraph{\textbf{ACT-R model.}}
The ACT-R model maintains jug states in the imaginal buffer. It includes rules for \texttt{fill}, \texttt{pour}, \texttt{empty}, and \texttt{reset}. A rule fires only if it avoids three conditions: (a) both jugs are empty, (b) both jugs are full, or (c) the next state would repeat a previous one. If all valid moves are exhausted, a reset rule clears the state. Because ACT-R does not evaluate long-term outcomes, the model may enter loops or redundant states unless additional planning mechanisms are introduced.
\section{Analysis}

Although we did not conduct automated formal verification, we translated our three cognitive models—Hungry-Thirsty, Fan Effect, and Water Jug—into state machine representations compatible with the nuXmv model checker. These translations were performed manually and served as a proof-of-concept for modeling cognitive architectures in a verification-friendly format. By encoding the control logic and memory updates in a declarative temporal modeling language, we were able to critically examine the structure and behavior of cognitive models across architectures and identify their verification-relevant features.

\subsection{Manual Translation Process}

Each ACT-R and Soar model was translated into a finite-state specification using nuXmv’s module-based language. For Soar, we constructed a top-level control module with variables representing system state, operator application, and internal memory flags (e.g., \texttt{state\_hungry}, \texttt{state\_thirsty}, \texttt{state\_operator\_name}). We introduced a `propose-apply` cycle that mimics Soar’s production rule sequencing. Each operator transitioned the model between cognitive states, and condition-action logic was encoded as transition rules. For ACT-R, we defined goal and buffer variables (\texttt{goal}, \texttt{goal\_hungry}, \texttt{goal\_thirsty}) and modeled the effects of production rules using guarded state transitions. Simplified representations of ACT-R’s buffer updates allowed us to abstract away subsymbolic components such as utility learning and activation dynamics, focusing instead on the symbolic flow of behavior.

\subsection{Observations from Modeling}

The translation process revealed several important insights:

\begin{itemize}
    \item \textbf{Explicitness of Control Flow:} Modeling in nuXmv forced the specification of state transitions and rule applicability in a highly explicit and deterministic manner. While Soar’s operators aligned well with this structure, ACT-R’s use of parallel buffer operations and utility-based selection required careful abstraction to fit a state machine model.
    
    \item \textbf{CMoC as an Intermediate Formalism:} Using the CMoC as a guide allowed us to identify architectural parallels between ACT-R and Soar. Constructs such as buffer updates, goal management, and conditional rule execution could be expressed in a common representational vocabulary, streamlining translation.

    \item \textbf{Simplification of Complex Features:} We deliberately omitted architectural features that are difficult to represent in purely symbolic systems (e.g., ACT-R’s stochastic retrieval, Soar’s chunking mechanisms). This simplification allowed for a cleaner mapping but also highlighted the need for future work in modeling probabilistic and adaptive behavior. While these omissions were necessary to demonstrate feasibility, they also limit external validity and will need to be addressed in future extensions.

    \item \textbf{Toolchain Reflection:} While manual translation was feasible for small models, the process was tedious and error-prone. This underscores the need for an automated compiler or modeling language that bridges cognitive architectures and formal verification environments, especially if larger or more dynamic models are to be verified.
\end{itemize}

Our manual translation serves as a stepping stone toward a future pipeline where formal properties can be specified and verified automatically. The current work demonstrates that both ACT-R and Soar models can be encoded in a verification-compatible format, and that CMoC provides a conceptual backbone for making this translation systematic and reusable.

\section{Conclusion}
This work demonstrates that formal verification methods—particularly symbolic model checking—can be meaningfully integrated with symbolic cognitive modeling through the CMoC. By developing and translating models into state machine representations, we showed how cognitive behaviors in ACT-R and Soar can be systematically prepared for analysis of correctness and termination using tools like \textit{nuXmv}.
Our current work is limited to three small-scale models and does not cover the full range of cognitive functions (e.g., learning, perceptual-motor integration). Moreover, while the CMoC abstraction is effective, it may require specialized handling during translation. In addition, our reliance on manual translation may introduce bias or artifacts, underscoring the importance of an automated compiler in future work.

\subsection{Future Work}

In the next phase of this research, we will develop a compiler that automates the transformation of models specified in the CMoC into executable code and formal verification models. This effort will include the design of a high-level modeling language that reflects the structure and semantics of CMoC components (perhaps similar to Herbal or to previous high level behavior representation languages), enabling more intuitive and architecture-neutral model specification. The compiler will support automatic translation into ACT-R and Soar representations for simulation, as well as into nuXmv-compatible formats for formal analysis. In parallel, we plan to extend the modeling framework to accommodate time-sensitive behaviors, such as response latencies and scheduling constraints, along with mechanisms for learning, including chunking and reinforcement learning. Our long-term objective is to scale the system to model complex, real-world cognitive tasks and to incorporate empirical data for validation, thereby enabling a unified, verifiable approach to cognitive modeling that bridges symbolic architectures and formal methods. 
\bibliographystyle{splncs04}
\bibliography{refs}
\end{document}